\documentclass[showpacs,twocolumn,prl,aps]{revtex4}
\usepackage{epsfig}
\usepackage{amsmath}
\usepackage{amssymb}
\begin{document}
\title{The Spin-SAF transition in $\rm NaV_2O_5$ induced by
       spin-pseudospin coupling}

\author{Claudius Gros and Gennady Y. Chitov} 

\affiliation{Fakult\"at 7, Theoretische Physik,
 University of the Saarland,
66041 Saarbr\"ucken, Germany.}

\date{\today}

\begin{abstract}
We present microscopic estimates for the 
spin-spin and spin-speudospin interactions
of the quarter-filled ladder compound
$\rm NaV_2O_5$, obtained by exactly diagonalizing
appropriate clusters of the underlying generalized
Hubbard Hamiltonian. We present evidence for a substantial
interladder spin-pseudospin interaction term which
would allow simultaneously for the superantiferroelectric
(SAF) charge (pseudospin) ordering and spin dimerization.
We discuss the values of the coupling constants appropriate
for $\rm NaV_2O_5$ and deduce the absence of a soft
antiferroelectric mode.
\end{abstract}

\maketitle

{\em Introduction}--
The insulating transition-metal compound
$\rm NaV_2O_5$ is, until now, the only
known quarter-filled symmetric ladder compound \cite{Lem03},
the closely related $\rm LiV_2O_5$ has asymmetric ladders
\cite{Val01}. It has a single electron per rung which
is located in the intra-rung $V^{4.5}-V^{4.5}$ 
bonding orbital, forming 
spin-1/2 Heisenberg chains in the high-temperature 
state \cite{Iso96}. $\rm NaV_2O_5$ has a dual
phase transition at $T_c=34\,{\rm K}$. Below this
temperature two things happen: 
A spin gap opens \cite{Fuj97}, like in a spin-Peierls
system and a crystallographic lattice distortion occurs
which leads to a charge disproportionation
$V^{4.5-\delta}-V^{4.5+\delta}$ on the rungs, alternating
along the ladder direction \cite{Most99,Most02,Gro99,Sma02}.
Note also that in the literature on $\rm NaV_2O_5$
its charge order is called the ``zig-zag phase", what characterizes
the antiferroelectric order in a single ladder only. In fact, the
two-dimensional long-range charge order in $\rm NaV_2O_5$ is
the super-antiferroelectric (SAF) \cite{Chi04}, and
we proposed to call the transition in $\rm NaV_2O_5$ the spin-SAF
transition. To develop an understanding of this
fascinating phase transition is an important matter,
as it may serve as a model for other systems with
coupled spin and orbital degrees of freedom\cite{Kug82}.

Basing ourselves on the earlier suggestion of
Mostovoy and Khomskii\cite{Most99,Most02} that a bilinear coupling between the
charge and spin degrees of freedom, similar to the spin-phonon coupling
in spin-Peierls systems, may be responsible for the transition
in $\rm NaV_2O_5$, we have proposed the theory of the spin-SAF
transition in the coupled spin-pseudospin model\cite{Chi03,Chi04,GCHS},
which we believe can explain the situation in that compound.
In our theory the simultaneous appearance of the SAF charge charge order
and of spin dimerization is driven by a single interladder spin-pseudospin
coupling.
Here we present results from a
microscopic study for the strength of
the coupling parameters necessary to understand the
spin-SAF phase transition in $\rm NaV_2O_5$.

\begin{figure}
\epsfig{file=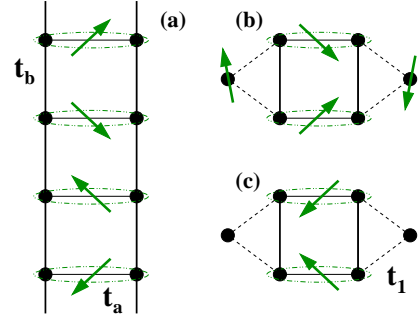,width=0.3\textwidth,angle=0}
\caption{The clusters used for the exact-diagonalization study.
(a) A 2x4 ladder with four electrons with are dominantly located
on the V-V bonding orbital, indicated symbolically by the dashed
ellipses.
(b) A six-site cluster with
two rungs and two sites of the respective adjacent ladders,
with four electrons. (c) Same as (b) but with two electrons.
        }
\label{fig_clusters}\end{figure}

{\em Microscopic Hamiltonian}--
The quarter-filled ladder compound $\rm NaV_2O_5$
is described microscopically by a generalized Hubbard model
\cite{Smo98} with
hopping matrix-elements $t_a$ and $t_b$ across the rung of
the ladder (the crystallographic a-direction) and
along the leg of the ladder respectively (the b-direction),
compare Fig.\ \ref{fig_clusters}, in addition to an interladder
hopping matrix element $t_1$. The Coulomb interaction between
the electrons gives rise to the onsite repulsion $U$, and the
intraladder n.n.\ matrix elements $V_a$ (rung) and $V_b$
(leg). Here we neglect the possible intraladder diagonal
repulsion $V_d$ and further interladder matrix elements.

{\em Pseudospin Hamiltonian}--
$\rm NaV_2O_5$ is an insulator and the low-energy
excitations of the microscopic generalized Hubbard 
model can therefore be mapped in perturbation theory
\cite{Most99,Deb00}, for small $t_b/U$ and small $t_b/V_a$,
to a spin-pseudospin Hamiltonian. Here we consider it in the form
$H=H_T+H_{S}+H_{ST}$, with
\begin{eqnarray}
\label{H_T}
H_T&=& 2t_a \sum_{n,m} T^x_{n,m} + \frac12 g
\sum_{m,n} T^z_{n,m} T^z_{n+1,m}\\
\label{H_S}
H_S&=& J_1 \sum_{n,m} {\bf S}_{n,m}\cdot{\bf S}_{n+1,m} \\
\label{H_ST}
H_{ST}&=&\varepsilon \sum_{n,m} {\bf S}_{n,m}\cdot{\bf S}_{n+1,m}
 \big( T^z_{n,m+1}- T^z_{n,m-1}\big)
\end{eqnarray}
where the spin/pseudospin operators obey the usual spin-algebra, e.g.\
$[T^\alpha,T^\beta]=i\epsilon_{\alpha,\beta,\gamma} T^\gamma$. The
sites indices $(n,m)$ count rungs/ladders.
In the high-temperature phase of
$\rm NaV_2O_5$ the electrons are delocalised on the rungs
with $\langle T_{n,m}^x\rangle\approx -1/2$. in the low-temperature phase
a finite-degree of charge ordering occurs with
$\langle T_{n,m}^z\rangle \propto (-1)^n$.

In (\ref{H_ST}) we have neglected further intraladder
spin-orbital coupling terms \cite{Deb00} which are not critically
relevant for the physics of the spin-SAF phase transition
in $\rm NaV_2O_5$ \cite{Chi03, Chi04}. The magnitude of the
spin-pseudospin coupling term $\varepsilon$ in (\ref{H_ST})
has not yet been estimated in perturbation theory, it
is $\sim t_bt_1^2$.

Recently we have shown \cite{Chi03},
that the spin- pseudospin coupling term
$H_{ST}\sim \varepsilon$ could at the same time drive
the observed phase transition in $\rm NaV_2O_5$ and 
lead to the observed opening of a gap in the spin-excitation
spectrum via an alternation of the effective spin-spin
coupling 
\begin{equation}
J_{n,m}^{\rm eff}\ =\
J_1+\varepsilon\, \langle T^z_{n,m+1}- T^z_{n,m-1} \rangle
\label{J_eff}
\end{equation}
along the ladder.

\begin{figure}
\epsfig{file=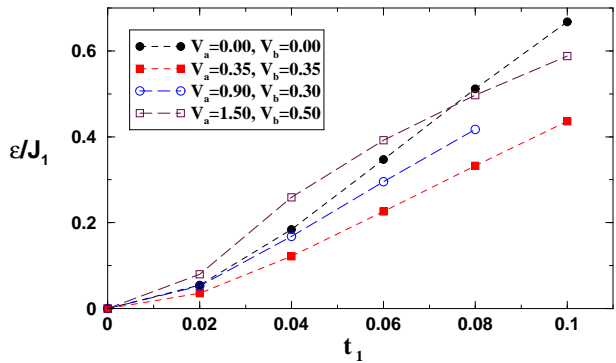,width=0.45\textwidth,angle=0}
\caption{Estimates of the interladder spin-pseudospin
coupling $\epsilon$, in units of the estimated intraladder
spin-exchange $J_1$, for $U=3$, $t_a=0.35$, $t_b=0.15$ by
comparing the respective triplet-excitation energies of the
six-site clusters (b) and (c) in Fig.\ \ref{fig_clusters}.
        }
\label{fig_epsilon}\end{figure}

{\em Spin-pseudospin coupling}--
The exchange coupling $J_1$ in (\ref{H_S})
can be estimated
by diagonalizing small clusters of $H$ 
and comparing the energy splitting between the
ground-state singlet and the first excited triplet
with the corresponding energy gap of the respective
finite-size spin model.

The inter-ladder spin-pseudospin coupling $\varepsilon$
can be determined using (\ref{J_eff}):
First we calculate the singlet-triplet 
energy gaps $\Delta E_t(N)$
of the
six-site clusters illustrated in Fig.\ (\ref{fig_clusters})
(b) and (c) respectively, i.e.\ with $N=4,2$
electrons. The position of the spins inside the
clusters shown in Fig.\ (\ref{fig_clusters}) 
illustrate typical spin-configurations realized
for the parameters relevant for $\rm NaV_2O_5$.
$\Delta E_t(N)$ therefore correspond to the
triplet-energy gaps of a system 
$J^{\rm eff}{\bf S}_1\cdot{\bf S}_2$
with two spins ${\bf S}_1$ and $ {\bf S}_2$
located on two n.n.\ rungs of the ladder, which
is just $J^{\rm eff}$. For the six-site cluster
with $N=4,2$ electrons we have approximatively

$$
\langle T^z_{n,m+1}- T^z_{n,m-1} \rangle
\ \approx \
\left\{\begin{array}{rc}
       1\quad & N=2\\
      -1\quad & N=4\\
       \end{array}\right.
$$
Using (\ref{J_eff}) we then find
\begin{equation}
\varepsilon\ = \ (\Delta E_t(2)
                 -\Delta E_t(4) )/2
\label{epsilon}
\end{equation}
for the spin-pseudospin coupling $\varepsilon$.
The results are presented in Fig.\ (\ref{fig_epsilon}).

\begin{figure}
\epsfig{file=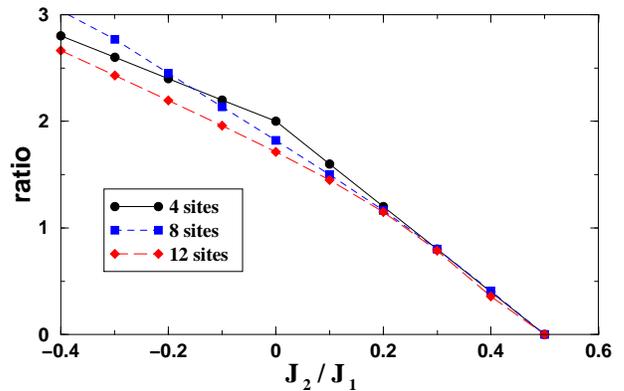,width=0.45\textwidth,angle=0}
\caption{Ratio $\Delta_s/\Delta_t$ of the singlet ($\Delta_s$)
and the triplet ($\Delta_t$) gap of spin-1/2 clusters with
periodic boundary conditions and  n.n./n.n.n.\ couplings $J_1$
and $J_2$ respectively. For $J_2/J_1>0.24$ this ratio 
extrapolates to zero in the thermodynamic limit.
        }
\label{fig_ratio}\end{figure}
{\em Parameters for $\rm NaV_2O_5$}--
The Coulomb repulsion $U$ has been estimated 
\cite{Smo98,Yar00} to
be $U\approx (3-4.1)\,{\rm eV}$ for $\rm NaV_2O_5$.
Its exact value is not critical and we use here
$U=3.0\,{\rm eV}$, all units throughout this papers are in
eV. We use $t_a=0.35\,{\rm eV}$ for the intra-rung hopping 
$t_a$, close to the LDA-estimate \cite{Smo98,Val01}
of $0.38\,{\rm eV}$. The inter-rung hopping $t_b$ along
the leg of the ladder is \cite{Smo98,Yar00} approximately
$0.15\,{\rm eV}-0.175\,{\rm eV}$.

The absolute magnitude of the
interladder hopping matrix element $t_1$ has
been difficult to determine from first principles.
It is relatively small \cite{Smo98,Yar00,Val01,Sua00} 
with a magnitude up to the estimated 
total interladder coupling of the order 
of $0.06\,{\rm eV}$.
Additionally, one has attributed
the smallness of the frustrating 
inter-ladder spin-spin coupling in the
high-temperature state \cite{Gro99,Sua00,Hor98}
to a mutual cancellation 
of antiferromagnetic and ferromagnetic
contributions due to intermediate singlet 
and triplet states \cite{Deb00,Hor98}.
The substantial size of $\varepsilon$ found
in our study and presented in Fig.\ \ref{fig_epsilon}
is then in qualitative agreement with the finding of
Yaresko {\it et al.} of
a substantial ferromagnetic
interladder spin-spin coupling, within a spin-resolved 
density-functional study of the fully
charge-ordered state \cite{Yar00}. We note that the
true amount of charge disproportionation
$\langle T^z_{n,m+1}- T^z_{n,m-1} \rangle$
in $\rm NaV_2O_5$ will be substantially reduced
from unity and with a corresponding reduction
of the size of the intraladder dimerization via
(\ref{J_eff}).

The size of the inter-size Coulomb repulsion
matrix elements $V_a$ and $V_b$ have been estimated
to be up to half an eV \cite{Deb00}, with a reduction
from the respective bare values due to screening.
The Ising-like pseudospin coupling-constant $g$
in (\ref{H_T}) is given by $g=4(V_b-V_d)$, where
$V_d$ the diagonal Coulomb repulsion
matrix element in between two rungs \cite{Deb00}.
Due to geometry
we have roughly $V_a\approx V_b \approx \sqrt 2 V_d$.
In order to avoid a proliferation of parameters we
have set here $V_d=0$ and examined two cases, see
Fig. \ref{fig_epsilon}:
$V_b=V_a$ and $V_b=V_a/3$. The latter case simulates
the reduction of $g\approx 4(V_b-V_b/\sqrt2)\approx 4V_a/3$
by $V_d$.

\begin{figure}
\epsfig{file=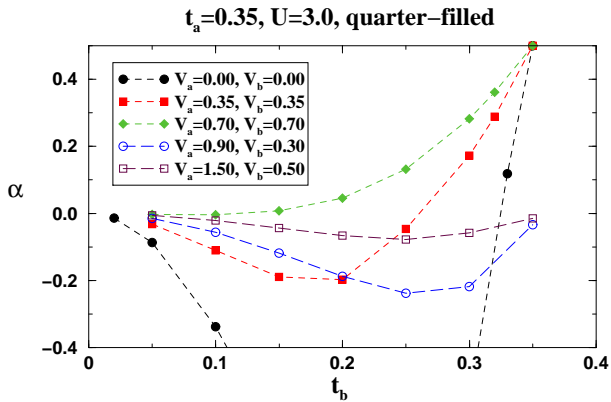,width=0.45\textwidth,angle=0}
\caption{The frustration parameter $\alpha=J_2/J_1$ for the 
n.n./n.n.n.\ couplings $J_1$ and $J_2$ along the ladder,
estimated by comparing $\Delta_s/\Delta_t$ for the 4-site
Heisenberg ladder (see Fig.\ \ref{fig_ratio}) with the results
obtained for  the $2\times4$ ladder 
(see Fig.\ \ref{fig_clusters}(a)) for $U=3$, $t_a=0.35$ and
various $V_a$ and $V_b$..
For $\rm NaV_2O_5$ the hopping $t_b$ along the leg is about
$t_b\approx0.175$.
        }
\label{fig_alpha}\end{figure}

{\em Intraladder frustration}--
Vojta {\it et al.} \cite{Voj01} 
have studied recently the possibility
of a intraladder spin gap formation for extended
quarter filled Hubbard models on a single two-leg ladder,
similar to the one studied here. They find a spontaneous
spin gap formation by DMRG for $t_b>t_a$. This value
for $t_b$ is outside of the parameter range relevant for
$\rm NaV_2O_5$ but of general interest. Here we present
a novel method to analyze finite-size data which allows
to obtain directly the microscopic frustration parameter
$\alpha=J_2/J_1$, where $J_2$ is the n.n.n.\ spin-coupling.

The frustrated spin-1/2 Heisenberg chain spontaneously
dimerizes for $\alpha>\alpha_c$ and the critical 
$\alpha_c\approx0.24$ \cite{OkNom}
can be accurately determined
by the crossing of the singlet excitation gap
$\Delta E_s(L)$ with the triplet excitation gap
$\Delta E_t(L)$ in chains with length $L$. For
$\alpha<\alpha_c$ we have $\Delta E_t(L)<\Delta E_s(L)$,
for $\alpha>\alpha_c$ the other way around.

In Fig.\ \ref{fig_ratio} we plot the ratio
$\Delta E_s(L)/\Delta E_t(L)$ for some small
Heisenberg chains of length $L=4,8,12$, as a function
of $J_2/J_1$. The Majumdar-Gosh state is exactly realized
for $\alpha=0.5$.
For $L=4$ there are only half as many
$J_2$-bonds than $J_1$-bonds and the Majumdar-Gosh state
is realized for $J_2=J_1$, we have therefore plotted
the data in Fig. \ \ref{fig_ratio} for $L=4$
as a function of $(J_2/2)/J_1$. 

The finite-size scaling
properties of $\Delta E_s(L)/\Delta E_t(L)$ are here,
however, unimportant. Here we point out that the
ratio $\Delta E_s(L)/\Delta E_t(L)$ is characteristic
for a given $L$. By calculating this ratio
for the four-rung quarter-filled ladder illustrated
in Fig.\ \ref{fig_clusters} we can therefor accurately
determine the degree of intraladder frustration.
The results are presented in Fig.\ \ref{fig_alpha}.
We confirm that the system spontaneously dimerizes
for $t_b=t_a=0.35$, for the parameters $V_a=V_b$
considered by Vojta {\it et al.} \cite{Voj01}.
We also find a tendency towards an anti-frustration,
i.e.\ a ferromagnetic $J_2$ for intermediate
values of $J_2$, indicating the absence of
frustration for $\rm NaV_2O_5$. In the limit
$V_a=0=V_b$ we find a very large and negative
$J_2$ indicating long-range interactions. 
Here the analysis breaks down, as it is valid only
for small to moderate values of $J_2$.

\begin{figure}
\epsfig{file=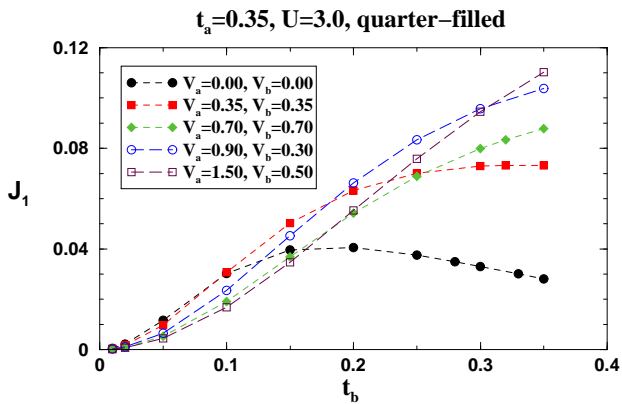,width=0.45\textwidth,angle=0}
\caption{Estimates for the n.n.\ spin-coupling $J_1$ along
the ladder, see Eq.\ (\ref{H_S}) from exact diagonalization
of the $2\times4$ ladder, see Fig.\ \ref{fig_clusters}(a).
The parameters are for $U=3$, $t_a=0.35$ and various
values for $V_a$ and $V_b$.
For $\rm NaV_2O_5$ the hopping $t_b$ along the leg is about
$t_b\approx0.175$.
        }
\label{fig_J1}\end{figure}

Finally we present in Fig.\ \ref{fig_J1} our
estimates for the n.n.\ intraladder spin coupling
$J_1$, obtained by comparing the triplet-gap
of the four-rung ladder (see Fig.\ \ref{fig_clusters})
with that of the four-site Heisenberg chain.
For $t_b\approx0.15-0.17$, the range relevant for
$\rm NaV_2O_5$ we find very reasonable value
$J_1=0.04-0.06$, close to the experimental value \cite{Iso96}
of $560\, {\rm K} = 0.048\,{\rm eV}$, which are
far more accurate than those by simple
perturbation theory \cite{Smo98}.

{\em Critical temperature}--
From the theory of the spin-SAF transition in the coupled spin-pseudospin
model \cite{Chi03,GCHS} we obtain the following equation for the critical
temperature at the Ising couplings $g< g_{\circ}$
(i.e., at the couplings where the pure transverse Ising model (\ref{H_T})
is always disordered)
\begin{equation}
\label{Tcimpl}
\eta(T_c/J_1)= {J_1 (g- g_{\circ})}/{4 \varepsilon^2},
\end{equation}
where $\eta(x)$ is the dimerization susceptibility of the dimerized
Heisenberg $XXX$ chain with the Hamiltonian
$
H_{\textrm{xxx}}=\sum_{n}
J_1(1+(-1)^n \Delta) {\textbf S}_{n}{\textbf S}_{n+1}
$ in the limit $\Delta \to 0$, and $g_{\circ}=4 t_a$ is the mean-field
approximation for the quantum critical
point of the transverse Ising model (\ref{H_T}).
The leading term of dimerization susceptibility,
as follows from the direct numerical (DMRG) calculations of Kl\"umper and
co-workers\cite{Klump}, can be reasonably approximated by the bosonization
result of Cross and Fisher \cite{CF79}, i.e., $\eta(x) \approx a_{\circ}/x$,
where we take $a_{\circ} \approx 0.26$. \cite{noteLog}
Using the above approximation for $\eta(x)$, we obtain the
critical temperature
\begin{equation}
\label{Tcas}
T_c  \approx
4 a_{\circ} \varepsilon^2/(g_{\circ}-g)
\end{equation}
as a function of the couplings.

{\em Parameters for $\rm NaV_2O_5$}--
Taking the experimental $T_c=34\,{\rm K}=2.93\,{\rm meV}$
and a spin-pseudospin coupling of about 
$\varepsilon\approx 0.4J_1=19.3\,{\rm meV}$
we find from Eq.\ (\ref{Tcas}) that
$g_\circ-g\approx 0.132\,{\rm eV}$, or
$g/g_\circ  \approx 0.91$.
This estimates for the intrachain Ising coupling implies
that $\rm NaV_2O_5$ is on the disordered side,
relatively close to the mean-field
quantum critical point of the transverse Ising model.
The standard RPA \cite{Most02,Blinc74} for the transverse
Ising model gives the following
spectrum of the pseudospin (charge) excitations
$
E_{\bf q}=\sqrt{2t_a[2t_a+m_x g({\bf q})]}
$,
where the pseudospin average $m_x=\tanh(t_a/T)/2 \approx 1/2$
at $T \ll t_a$
and $g({\bf q})$ is the Fourier transform of the Ising coupling.
 $E_{\bf q}$ corresponds to the pole of the retarded pseudospin-pseudospin
 correlation function $\langle T^z T^z \rangle(\omega, {\bf q})
 \equiv D_{zz}(\omega, {\bf q})
 =2t_a m_x/(\omega^2-E_{\bf q}^2)$ \cite{Most02}. 
In the function $g({\bf q})$ found earlier on the effective
2D lattice \cite{Chi04} we retain only the largest intraladder coupling g,
so ($g <g_{\circ}$)
\begin{equation}
\label{EqGaps}
E_{0/{\bf q}_{\rm  SAF}}=2t_a \sqrt{1 \pm g/g_{\circ} }~.
\end{equation}
With the above estimates for the couplings we get
$E_0 \approx 0.97\,{\rm eV}\approx 7800~cm^{-1}$ which agrees well with
the observed peak of the optical conductivity (${\bf q} \approx 0$)
\cite{Damascelli00}. At ${\bf q}={\bf q}_{\rm SAF}$ corresponding
to the SAF ordering wave-vector (${\bf q}_{ \rm SAF}=(0,\pi)$ on the
effective 2D square lattice \cite{Chi04}) we get
$E_{{\bf q}_{\rm  SAF}} \approx 0.21\,{\rm eV}\approx 1700~cm^{-1}$.

{\em Discussion}--
From the Cross-Fisher RPA for the coupled spin-phonon system \cite{CF79},
Gros {\it et al} have shown \cite{Gros_Werner}
that a soft mode (i.e., a pole in the phonon-phonon correlation function)
occurs at a spin-Peierls transition only
if the energy of the involved phonon mode $\Omega$ is small,
namely, $\Omega<2.2\,T_{SP}$. The RPA for the coupled spin-pseudospin
model renormalizes the pseudospin correlation function as
$D_{zz,R}^{-1}=D_{zz}^{-1}-\Pi$ with $\Pi=-4\varepsilon^2 \eta$,
resulting in Eq.~(\ref{Tcimpl}) for the critical temperature of the
spin-SAF transition and the same condition for the occurrence of a
pole of $D_{zz,R}$. A soft antiferroelectric mode is therefore absent
in $\rm NaV_2O_5$, since
$E_{{\bf q}_{\rm  SAF}} \approx 210\,{\rm meV}>2.2\,T_c=6.4\,{\rm meV}$.

The energy $2E_{{\bf q}_{\rm  SAF}} \approx 3400~cm^{-1}$
is quite close to the position of the broad peak near $4000~cm^{-1}$
observed in the optical absorption experiments at the room temperature
\cite{Damascelli00} (see also \cite{Most02} for more detailed
data in the frequency range concerned).
This suggests that this peak can be attributed to the two-particle
(pseudospin) excitations with the ordering wave-vectors.

{\em Conclusions}-- We have proposed that the spin-SAF transition
is of novel kind, in the sense that it is driven directly by
the coupling in between spin and orbital degrees of freedoms.
We have presented support for this scenario from numerical
estimates for the coupling parameters and proposed an mechanism
for a hitherto unexplained feature in the infrared spectrum.
These studies might be of relevance for other materials with
strong spin-orbital couplings.

%

\end{document}